\documentclass[reviewt]{elsarticle}

\usepackage{siunitx}
\journal{NIM-A}

\begin{document}

\begin{frontmatter}

\title{Status report of the ESCULAP project at Orsay: External injection of low energy electrons in a Plasma.}

\author[CLUPS]{Elsa Baynard}
\author[LAL]{Christelle Bruni}
\author[LAL]{Kevin Cassou}
\author[LAL]{Vincent Chaumat}
\author[LAL]{Nicolas Delerue\corref{mycorrespondingauthor}}
\cortext[mycorrespondingauthor]{Corresponding author}
\ead{delerue@lal.in2p3.fr}
\author[LPGP]{Julien Demailly}
\author[LAL]{Denis Douillet}
\author[LAL]{Noureddine El Kamchi}
\author[LIDYL]{David Garzella}
\author[LPGP]{Olivier Guilbaud}
\author[LAL]{Stephane Jenzer}
\author[LPGP]{Sophie Kazamias}
\author[LAL]{Viacheslav Kubytskyi}
\author[LAL]{Pierre Lepercq}
\author[LPGP]{Bruno Lucas}
\author[LPGP]{Gilles Maynard}
\author[LPGP]{Olivier Neveu}
\author[CLUPS]{Moana Pittman}
\author[CLIO]{Rui Prazeres}
\author[LAL]{Harsh Purwar}
\author[LPGP]{David Ros}
\author[LAL]{Cynthia Vallerand}
\author[LAL]{Ke Wang}

\address[CLUPS]{CLUPS, Univ. Paris-Sud, Universit\'e Paris-Saclay, Orsay, France.}
\address[LAL]{LAL, Univ. Paris-Sud, CNRS/IN2P3, Universit\'e Paris-Saclay, Orsay, France.}
\address[LPGP]{Laboratoire de Physique des Gaz et des Plasmas, Univ. Paris-Sud, CNRS, Universit\'e Paris-Saclay, Orsay, France.}
\address[CLIO]{CLIO/LCP, Univ. Paris-Sud, CNRS, Universit\'e Paris-Saclay, Orsay, France.}
\address[LIDYL]{CEA/DRF/LIDYL, Universit\'e Paris-Saclay, Saclay, France.}

\begin{abstract}
The ESCULAP project aims at studying external injection of low energy (\SI{10}{MeV}) electrons in a plasma in the quasilinear
regime. This facility will use the photo injector PHIL and the high power laser LASERIX. We will give a status report of
the preliminary work on the facility and the status of the two machines. We will also present the results of simulations
showing the expected performances of the facility.

\end{abstract}

\begin{keyword}
External injection, Laser plasma acceleration, LPWA, gas cell, magnetic compression
\end{keyword}

\end{frontmatter}

%\linenumbers

\section{Overview of the ESCULAP project.}

The ESCULAP (ElectronS CoUrts pour l'Acc\'el\'eration Plasma)~\cite{Delerue:2016tcy} project aims at studying external injection of low energy (\SI{10}{MeV}) electrons in a plasma in the quasilinear regime. This experiment will use the photo injector PHIL~\cite{1748-0221-8-01-T01001} and the high power laser LASERIX~\cite{Guilbaud:15}. 

\begin{figure}[htbp]
  \centering
  \includegraphics[width=1.1\linewidth]{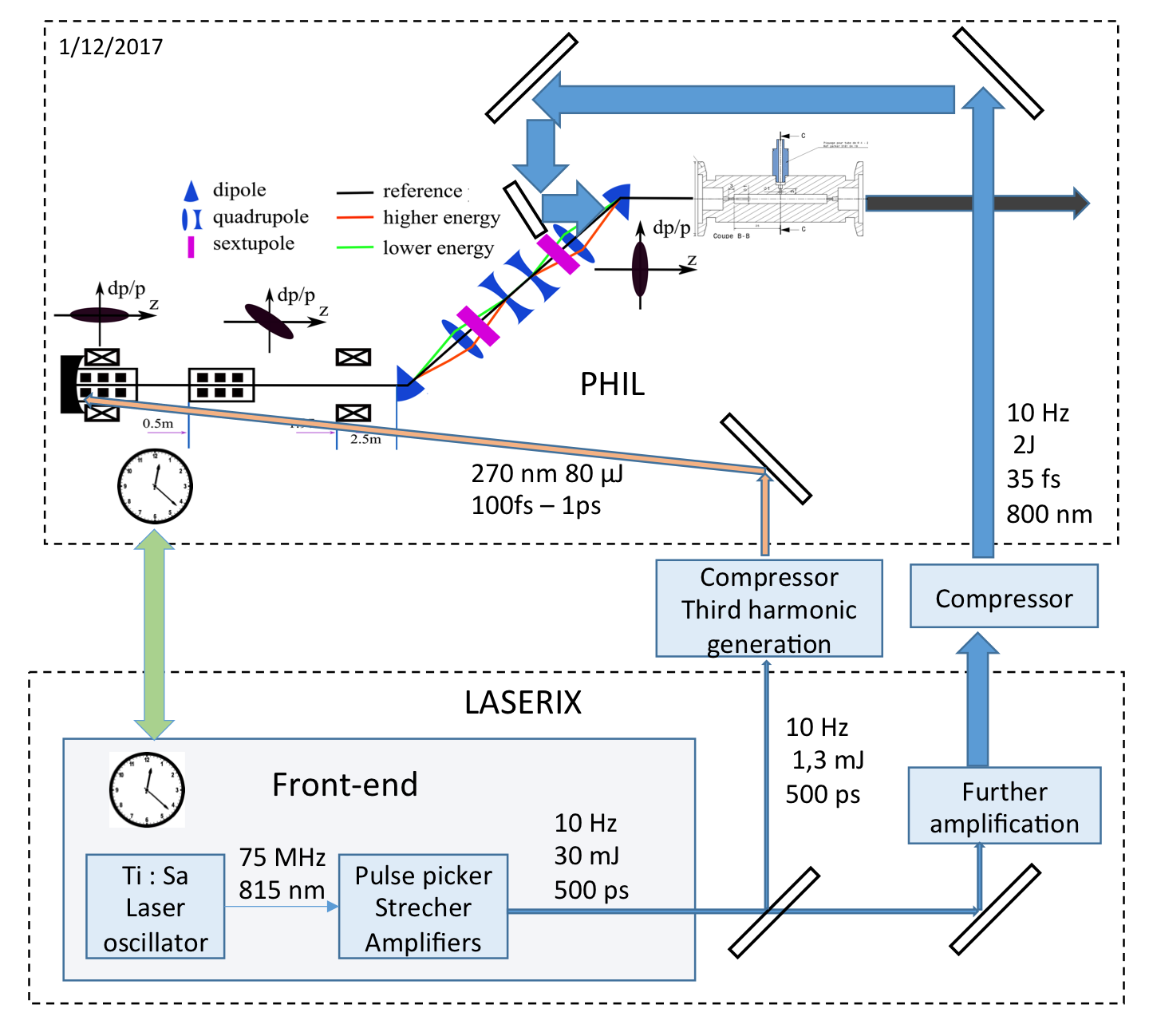}
  \caption{The proposed layout for the ESCULAP experiment: the PHIL and Laserix facilities are combined for this external injection laser-driven plasma acceleration experiment.}
  \label{fig:layout_ESCULAP}
\end{figure}

The proposed layout of the experiment is shown in figure~\ref{fig:layout_ESCULAP}. The electron bunches coming from PHIL are compressed in a compression chicane and then injected in a plasma cell where they are accelerated.
 In this paper,  section~\ref{sec:machines} will describe the two machines forming the project,  section~\ref{sec:qlinear}  will present the advantages of quasilinear acceleration, section~\ref{sec:synchro}  will describe the synchronization of the two machines,  section~\ref{sec:chicane}  will describe the compression chicane,  section~\ref{sec:cell}  will present the constraints on the gas cell and the final part of the paper will give an outlook of the project.

\section{PHIL and Laserix}
\label{sec:machines}
\subsection{PHIL}
PHIL is a \SI{5}{MeV} photo injector that has been described in~\cite{1748-0221-8-01-T01001}. With a copper photocathode it can produce electron bunches with a charge up to \SI{100}{pC} and an emittance of \SI{1}{mm.mrad} (at the cathode) however for this project it will be used to produce lower charges (\SI{10}{pC} for the first experimental campaign).

\subsubsection{PHIL upgrade}

PHIL will be upgraded so that it can produce electron bunches with an energy of \SI{9}{MeV}. A \SI{3}{GHz} 3-cell standing wave 
booster will be installed after the photo-injector. In this cavity an RF input power of \SI{3}{MW} will produce the \SI{50}{MV/m} accelerating
gradient required to increase the energy of the beam up to \SI{9}{MeV}.

\subsection{Laserix}
LASERIX is a \SI{50}{TW} Ti-Sapphir laser chain mainly dedicated to provide two XUV beamlines that can be either separated or combined together which is described in~\cite{Guilbaud:15}. It is part of the pettawatt system described in~\cite{Ple:07} and being re-installed at Universit\'e Paris-Sud. 
For the present purpose, the Ti-Sapphir laser chain delivers laser beam at \SI{10}{Hz} repetition rate, with an energy of about \SI{2}{J}  and \SI{40}{fs} pulse duration at a wavelength of \SI{810}{nm}. A sampling of \SI{1.3}{mJ} is frequency tripled to UV (\SI{270}{nm}) with a third harmonic generator and sent to the PHIL photocathode. The \SI{2}{J} energy beam will be focused in a plasma cell to create the plasma wave.

\section{Modelisation of the experiment}
\label{sec:qlinear}
An extensive effort has been done to simulate the experiment and is document in~\cite{Delerue:2016tcy} and~\cite{EAAC17_simulations}. The simulations results show that after acceleration the electrons should reach an energy of about~\SI{140}{MeV} with a trapping efficiency better than \SI{70}{\percent} for \SI{10}{pC} injected.

\section{Synchronisation of the high power laser LASERIX with the Photoinjector PHIL}
\label{sec:synchro}
As the two machines use independent clocks an heterodyne synchronisation system has been devised and tested. It has been presented in~\cite{Delerue:2017kxo}. It uses a mixer to generate an heterodyne signal from the PHIL \SI{3}{GHz} RF and the signal from a photodiode located inside the LASERIX oscillator. This heterodyne signal is then low-pass filtered and one zero crossing is used to generate the trigger signals required by PHIL and LASERIX. Measurements with an ultra-fast scope have shown that the jitter of this synchronization system is lower than~\SI{2}{ps} rms.

\section{Compression of the electron bunch}
\label{sec:chicane}

\begin{figure}[htbp]
  \centering
  \includegraphics[width=0.9\linewidth]{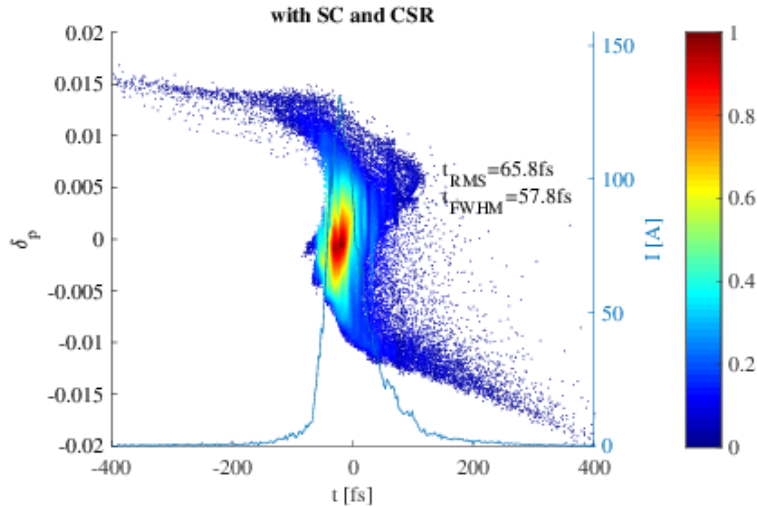}
  \caption{Phase space of the particles after the ESCULAP compression chicane as described in~\cite{EAAC17_compression}.}
  \label{fig:chicane}
  \end{figure}

To match the parameters of the electron beam with those required by the plasma a compression chicane is required. It will require at least two sets of quadrupoles and a pair of sextupoles to provide longitudinal compression but additional quadrupoles will be used for transverse compression. Simulations with ImpactT~\cite{PhysRevSTAB.9.044204} of the beam expected after the chicane are shown on figure~\ref{fig:chicane} and are described in detail in~\cite{EAAC17_compression}, together with the chicane. As we can see such chicane can compress the beam to much better than our target of~\SI{100}{fs} FWHM and in~\cite{EAAC17_compression} it is also shown that the beam can be focussed to better than \SI{150}{\micro m} FWHM in the transverse plane at the entrance of the plasma.

\section{Gas cell}
\label{sec:cell}

\begin{figure}[htbp]
  \centering
  \includegraphics[width=1.1\linewidth]{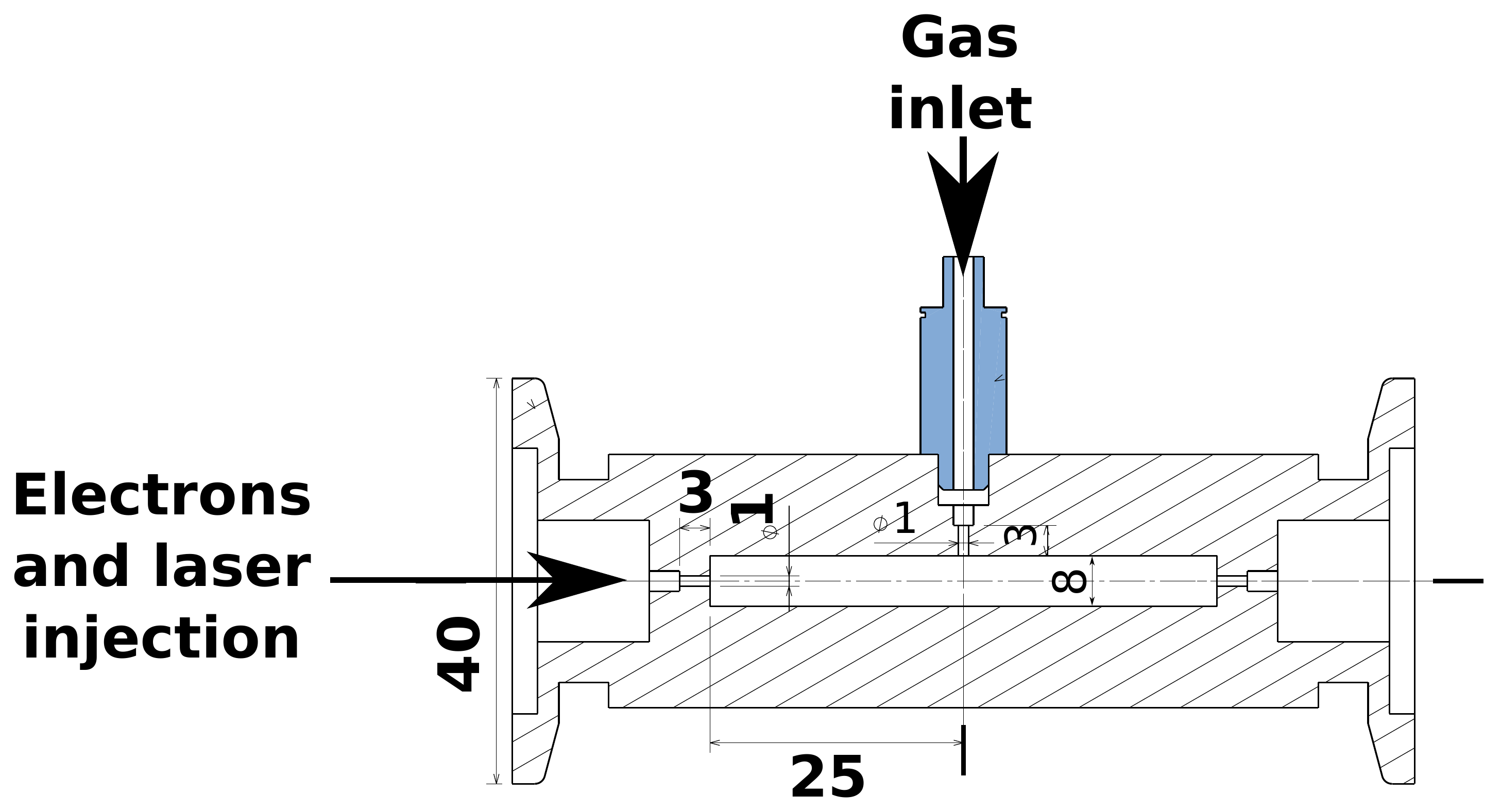}
  \caption{Preliminary design of the plasma cell considered for ESCULAP. }
  \label{fig:ESCULAP_cell}
  \end{figure}

The next challenge to be addressed is to be able to have a variable plasma density along the beam axis. Given the low energy at which we will perform the experiments this shaping will be rather important.  Several strategies are currently being investigated: either by shaping the channel in which the gas will propagate or by using different gas inlets at different pressures. A preliminary design of such a cell is shown in figure~\ref{fig:ESCULAP_cell}. This plasma cell has quartz windows on two sides to allow imaging of the plasma density along most of its length.

\section{Outlook}

The ESCULAP experiment will investigate laser-driven plasma acceleration by external injection in the quasilinear regime. It will be a unique facility in its energy injection range. We have demonstrated that the two facilities PHIL and LASERIX can be synchronized to better than~\SI{2}{ps}. We have also shown that the electrons could be focussed longitudinally to better than~\SI{100}{fs} and transversely to better than~\SI{150}{\micro m}. The design of the plasma cell is well under way.

\section{Acknowledgements}
The ESCULAP team is grateful to CNRS and Univ. Paris-Sud (Universit\'e Paris-Saclay) for their support.

\section*{References}

\bibliography{biblio}

\end{document}